\documentclass[aps,prl,twocolumn,amsmath,amssymb,superscriptaddress]{revtex4}

\newcommand{\bra}[1]{\langle#1|}
\newcommand{\ket}[1]{|#1\rangle}

\usepackage[pdftex]{graphicx}
\usepackage{mathrsfs}

\begin{document}

\bibliographystyle{apsrev}

\title{Time-resolved detection and mode-mismatch in a linear optics quantum gate}

\author{Peter P. Rohde}
\email[]{dr.rohde@gmail.com}
\homepage{http://peterrohde.wordpress.com}
\affiliation{Max-Planck Institute for the Science of Light, Erlangen, Germany}
\affiliation{Centre for Quantum Computation and Communication Technology, School of Mathematics and Physics, University of Queensland, Australia}

\author{Timothy C. Ralph}
\affiliation{Centre for Quantum Computation and Communication Technology, School of Mathematics and Physics, University of Queensland, Australia}

\date{\today}

\frenchspacing

\begin{abstract}
Linear optics is a promising candidate for the implementation of quantum information processing protocols. In such systems single photons are employed to represent qubits. In practice, single photons produced from different sources will not be perfectly temporally and frequency matched. Therefore understanding the effects of temporal and frequency mismatch is important for characterising the dynamics of the system. In this paper we discuss the effects of temporal and frequency mismatch, how they differ, and what their effect is upon a simple linear optics quantum gate. We show that temporal and frequency mismatch exhibit inherently different effects on the operation of the gate. We also consider the spectral effects of the photo-detectors, focusing on time-resolved detection, which we show has a strong impact on the operation of such protocols.
\end{abstract}

\maketitle

Linear optics quantum computing (LOQC) \cite{bib:KLM01} has emerged as a promising candidate for the implementation of quantum information processing \cite{bib:NielsenChuang00} protocols. LOQC protocols are essentially large interferometers, where single photons represent qubits. Typically the implementation of such protocols requires the indistinguishability of photons, such that the desired interference takes place. However, in practice photons will not be completely indistinguishable, which undermines the desired interference. Therefore, understanding the effects of photon distinguishability is important.

In this paper we investigate the effects of spectral and temporal mismatch on the operation of a simple linear optics (LO) quantum gate and especially focus on how the parameters characterising the photo-detectors influence such protocols. The new result is that we demonstrate that temporal and spectral mismatch have inherently different effects on the operation of LO protocols. Previous authors have investigated temporal mismatch in LO gates \cite{bib:RohdeRalph05, bib:RohdePryde05,bib:RohdeRalph06,bib:RohdeRalphMunro06}, as well as spectral mismatch in the context of a distributed quantum entanglement protocol based on LO \cite{bib:MetzBarrett08}. Here we build on these previous studies by reconciling these two effects. In particular, we consider the case where the detectors are able to resolve sub-wavepacket arrival times and how this additional timing information affects the dynamics of the system.

Legero \emph{et al.} \cite{bib:Legero03,bib:Legero05} made the observation that with time-resolved detectors novel `quantum beating' effects can be observed in a Hong-Ou-Mandel (HOM) interferometer \cite{bib:HOM87}. This arises when the detector response time is much smaller than the length of the interacting photons' temporal wavepackets. This phenomena has been experimentally demonstrated \cite{bib:Legero04}. Here we consider such effects in the context of an elementary LO quantum gate. We focus on a time-integrated detector model \cite{bib:RohdeRalph06b}, and examine the effects of both temporal and frequency mismatch on the operation of the gate. We observe that the gate can generally be implemented with high fidelity when the detector integration times are sufficiently small and the detectors click simultaneously, although this will come at the expense of gate success probability. Temporal mismatch generally results in monotonic deterioration of the fidelity of the gate, whereas frequency mismatch results in an oscillatory behaviour whereby perfect gate operation periodically arises.

Temporal and frequency mismatch arise naturally in many physical systems. For example, when coupling two independent photon sources into a quantum gate, perfect temporal overlap is difficult to achieve. Similarly, in some photon sources identical spectral structure between different sources is challenging. For example, with a system comprising an atom in a cavity, perfect control over the cavity frequency is not always possible. In parametric down-conversion (PDC) sources, independent sources will not exhibit perfectly identical phase-matching conditions, resulting in different spectral properties of the independently produced photons. Similarly, PDC sources based on triggering will not produce identical spectral structures owing to spectral imperfections in the triggering detectors. These effects are evident in many HOM-type experiments where results close to unit visibility are difficult to achieve, indicative that temporal, frequency and/or spatial mismatch are occurring.

\begin{figure}[!h]
\includegraphics[width=0.8\linewidth]{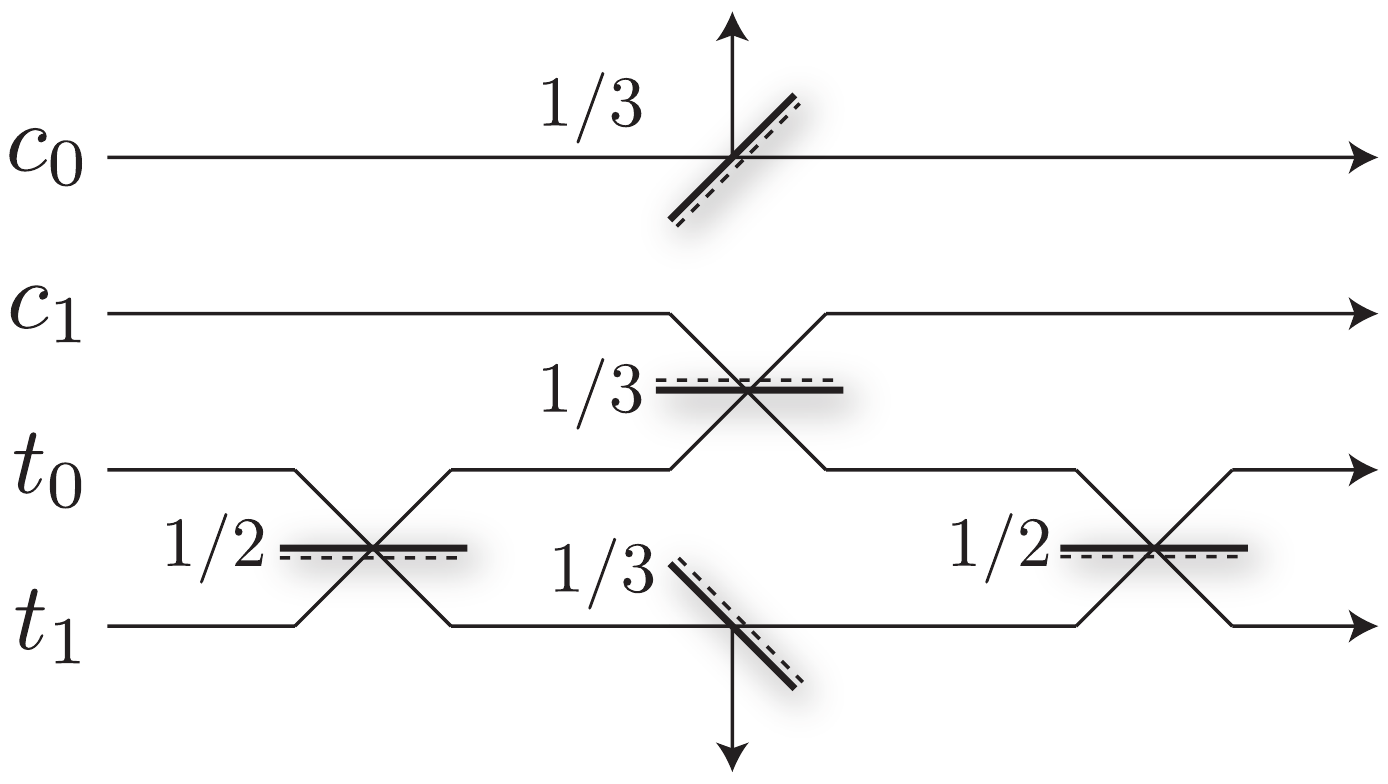}
\caption{The coincidence CNOT gate. There are two qubits $c$ and $t$, each encoded across two spatial modes. We post-select upon detecting exactly one photon in the $t$ outputs and exactly one photon in the $c$ outputs. Upon post-selection the gate implements the CNOT operation. The output modes are on the right, while the central two exiting modes are discarded. A $\pi$ phase-shift is induced upon reflection from the dotted sides of the beamsplitters.} \label{fig:CNOT}
\end{figure}

We will focus on the coincidence CNOT gate \cite{bib:Ralph02} shown in Fig. \ref{fig:CNOT}. This is a non-deterministic gate employing dual-rail encoding, which succeeds upon post-selection of exactly one photon in the control output and one photon in the target output. This gate was first experimentally demonstrated by O'Brien \emph{et al.} \cite{bib:OBrien03} and has since been used in many quantum circuits \cite{bib:Pryde05,bib:Kiesel05,bib:Lanyon09,bib:Politi09}. We have chosen this gate because it encompasses all the main interesting features of LO quantum gates: (1) the CNOT gate is ubiquitous in quantum information processing applications, and appears in many circuits and protocols, (2) the CNOT gate is a maximally entangling gate, (3) the coincidence CNOT gate contains both HOM and Mach-Zehnder type interference, and (4) the coincidence CNOT gate relies on two independently prepared photons, which lends itself to analysing the effects of photon distinguishability.
The CNOT gate can be characterised by a truth table, which defines the mapping from the input basis states to the output basis states. In the computational, $Z$, basis this is given by,
\begin{equation} \label{eq:ideal_CNOT}
\left( \begin{array}{cccc}
1 & 0 & 0 & 0 \\
0 & 1 & 0 & 0 \\
0 & 0 & 0 & 1 \\
0 & 0 & 1 & 0
\end{array} \right),
\end{equation}
where we have used the logical basis $\ket{00}$, $\ket{01}$, $\ket{10}$, $\ket{11}$ and the first qubit is the \emph{control} and the second the \emph{target}. To demonstrate the true quantum behaviour of the gate it is also necessary to verify the operation of the gate in a non-commuting basis, since a CNOT truth table may also exist for a classical XOR gate. For the subsequent study we verified that the gate also behaves equivalently when the input states are in the diagonal, $X$, basis.

We define the form of a single photon state,
\begin{equation} \label{eq:temporal_dist}
\int \psi(t) a(t)^\dag dt \ket{0},
\end{equation}
where $\psi(t)$ is the temporal distribution function of the photon and is related to the spectral distribution via a Fourier transform \footnote{Strictly the Fourier transform relation is only approximate as the frequency spectrum is restricted to positive frequencies. However, for typical pulse widths at optical frequencies this approximation is very good.}.

If a detector registers a count in a very short time interval $\delta$ around time $t_0$, such that \mbox{$\psi(t_0) \approx \psi(t_0 + \delta)$}, then the state of Eq. \ref{eq:temporal_dist} is effectively projected on to
\begin{equation}
\psi(t_0) a(t_0)^\dag \ket{0},
\end{equation}
where $|\psi(t_0)|^2$ can be regarded a probability density function. If the detector integrates over some range $t_w$, then the measurement probability is
\begin{equation}
p(t_0) = \int_{t_0}^{t_0 + t_w} |\psi(t)|^2 dt.
\end{equation}
In this study we will consider two different detector scenarios -- time-resolved and gated detection. In time-resolved detection the detector tells us the measurement time of the photon, up to an uncertainty given by the integration time. In general the detection times for multiple photons may differ. In gated detection our detector is only able to trigger in a very short, predetermined time window, in which case we will focus on the case where the gate times for different detectors are the same. This is equivalent to using time-resolved detection and post-selecting on events where the photons are measured at the same time. See Ref. \cite{bib:MetzBarrett08} for an alternate, more physically motivated, model for time-resolved detection.

The input state to the CNOT gate is
\begin{eqnarray} \label{eq:input_state}
\ket{\psi_{in}} &=& \lambda_{00} \int \psi_{c}(t) a(t)_{c_0}^\dag dt \int \psi_{t}(t') a(t')_{t_0}^\dag dt' \ket{0}  \nonumber \\
&+& \lambda_{01} \int \psi_{c}(t) a(t)_{c_0}^\dag dt \int \psi_{t}(t') a(t')_{t_1}^\dag dt' \ket{0} \nonumber \\
&+& \lambda_{10} \int \psi_{c}(t) a(t)_{c_1}^\dag dt \int \psi_{t}(t') a(t')_{t_0}^\dag dt' \ket{0} \nonumber \\
&+& \lambda_{11} \int \psi_{c}(t) a(t)_{c_1}^\dag dt \int \psi_{t}(t') a(t')_{t_1}^\dag dt' \ket{0}, \nonumber \\
\end{eqnarray}
where $c$ and $t$ denote the control and target qubits, and $0$ and $1$ denote the two spatial modes associated with each qubit. $\lambda$ denote the coefficients of the logical basis states. We choose the spectral distribution functions to be Gaussian distributions, where one of the photons is spectrally ($\omega$) and temporally ($\tau$) displaced relative to the other,
\begin{eqnarray} \label{eq:tdf}
\psi_c(t) &=& \sqrt[4]{\frac{2}{\pi }} e^{-t^2} \nonumber \\
\psi_t(t) &=& \sqrt[4]{\frac{2}{\pi }} e^{-i \omega  t} e^{-(t-\tau )^2}.
\end{eqnarray}
We have implicitly assumed the two control modes share the same temporal distribution function, as do the two target modes. This is a realistic assumption when the gate is demonstrated in isolation and each of the qubits emanate from a single photon source. However, this assumption need not hold when gates are cascaded.

Upon measurement, the probability distribution function for the truth table of the gate is of the form
\begin{equation}
p(t_c,t_t) = \left( \begin{array}{cccc}
\alpha_{t_c,t_t} & 0 & 0 & 0 \\
0 & \alpha_{t_c,t_t} & 0 & 0 \\
0 & 0 & \beta_{t_c,t_t} & \gamma_{t_c,t_t} \\
0 & 0 & \gamma_{t_c,t_t} & \beta_{t_c,t_t}
\end{array} \right),
\end{equation}
where $t_c$ and $t_t$ are the times at which we measure the control and target respectively. The parameters in the matrix are given by
\begin{eqnarray}
\alpha_{t_c,t_t} &=& \frac{2 e^{-2 \left(\left(\tau -t_c\right){}^2+t_t^2\right)}}{9 \pi }, \nonumber \\
\beta_{t_c,t_t} &=&\frac{2 \left | e^{-t_c^2-\left(\tau -t_t\right){}^2-i \omega  t_t}-e^{-\left(\tau -t_c\right){}^2-i \omega  t_c-t_t^2}\right |^2}{9 \pi }, \nonumber \\
\gamma_{t_c,t_t} &=& \frac{2 e^{-2 \left(t_c^2+\left(\tau -t_t\right){}^2\right)}}{9 \pi }.
\end{eqnarray}
These expressions are obtained by propagating the input state from Eq. \ref{eq:input_state} with the temporal distribution functions given in Eq. \ref{eq:tdf} through the network and performing ideal time-resolving measurements at the outputs, represented using projectors of the form $\ket{t_c}\bra{t_c}\otimes\ket{t_t}\bra{t_t}$. The interesting interference takes place at the central $1/3$ beamsplitter, which mixes the control and target qubits.

When there is no time and frequency shift in the target distribution (i.e. $\tau=0$ and $\omega=0$) this matrix reduces to the CNOT matrix, up to a constant factor. This is expected since in this case we are dealing with indistinguishable photons. Similarly, when both detectors click at the same time, \mbox{$t_t = t_c$}, we observe ideal operation since the detectors are unable to distinguish the two photons. More generally, when there is no temporal mismatch, $\tau=0$, and the detectors click at different times, we observe perfect gate operation when \mbox{$\omega (t_c - t_t) = 2\pi n$}, for integer $n$. When post-selecting the gate such that ideal operation can be achieved, the latter observation allows us to boost the success probability of the gate, since there are now multiple click times which result in ideal gate operation.

It can easily be verified that the above expressions for $\alpha$, $\beta$ and $\gamma$ are invariant under a common shift in centre frequency of the two photons. Thus in our subsequent results we ignore such global translations.

We characterise the operation of the gate using the \emph{similarity} - a fidelity measure for classical probability distributions - of the gate's truth table with the ideal CNOT truth table,
\begin{equation}
S = \frac{\left( \sum_{i,j} \sqrt{M_{i,j} M'_{i,j}}\right)^2}{\sum_{i,j}M_{i,j} \sum_{i,j}M_{i,j}'},
\end{equation}
where $M$ and $M'$ are the two truth tables being compared and $M_{i,j}$ is the element of the truth table in row $i$ and column $j$. Note that the similarity measure inherently renormalises the matrices, so matrices representing gates with different success probabilities can still be fairly compared.

For the gate in question we calculate
\begin{equation}
S = \frac{\left(e^{2 \tau  t_c}+e^{2 \tau  t_t}\right)^2}{4 \left(e^{4 \tau  t_c}+e^{4 \tau  t_t}-e^{2 \tau  \left(t_c+t_t\right)} \mathrm{cos}\left[\omega  \left(t_c-t_t\right)\right]\right)}.
\end{equation}
Two properties immediately follow from this expression. First, when $t_t=t_c$, $S=1$ and we have perfect gate operation, as noted above, because both detectors are clicking at the same time hence they cannot reveal any distinguishing information about the photons. Second, when both $\tau=0$ and $\omega=0$, $S=1$ since now the photons are completely indistinguishable, both temporally and spectrally, and perfect interference must take place.

These observations apply when we are dealing with perfect detectors that project onto infinitesimal temporal states. To calculate the overall truth table for some integration window, we define 
\begin{equation}
p(t_c,t_t) = \int_{t_c}^{t_c + t_w} \! \! \! \! \int_{t_t}^{t_t + t_w} \left( \begin{array}{cccc}
\alpha_{t,t'} & 0 & 0 & 0 \\
0 & \alpha_{t,t'} & 0 & 0 \\
0 & 0 & \beta_{t,t'} & \gamma_{t,t'} \\
0 & 0 & \gamma_{t,t'} & \beta_{t,t'}
\end{array} \right) dt\,dt'.
\end{equation}
Here the truth table consists of classical probabilities, so phase relations are ignored. Importantly, in the ideal CNOT gate from Eq. \ref{eq:ideal_CNOT} the success probability of the gate is independent of the input state. However, in the general case the different logical basis states are transformed with differing success probabilities. Thus the operation of the gate is biased as a function of $\alpha$, $\beta$ and $\gamma$. A lower bound on the success probability of the gate, across all basis states, is given by
\begin{eqnarray}
p_\mathrm{min} = \mathrm{min} \Big(\!\!\!\!\!\! && \int_{t_c}^{t_c + t_w} \! \! \! \! \int_{t_t}^{t_t + t_w} \alpha_{t,t'} \, dt  \, dt', \nonumber \\
&&\int_{t_c}^{t_c + t_w} \! \! \! \! \int_{t_t}^{t_t + t_w} \beta_{t,t'} + \gamma_{t,t'} \, dt \, dt' \Big).
\end{eqnarray}
It should be noted that as the integration window $t_w$ is reduced, the success probability of the gate drops. Indeed, with mismatched photons the success probability is further reduced by gating -- with mismatched photons, two photons are very unlikely to be found within the same narrow window. However, this is the price to pay for improved gate fidelity as we will discuss shortly. This is a major obstacle for experimentalists, who routinely employ gating techniques, which undermines the success probability of their gates. A demonstration of elementary optical quantum gates without the need for narrowband filtering would be a major step forward.

\begin{figure}[h!tb]
\includegraphics[width=\linewidth]{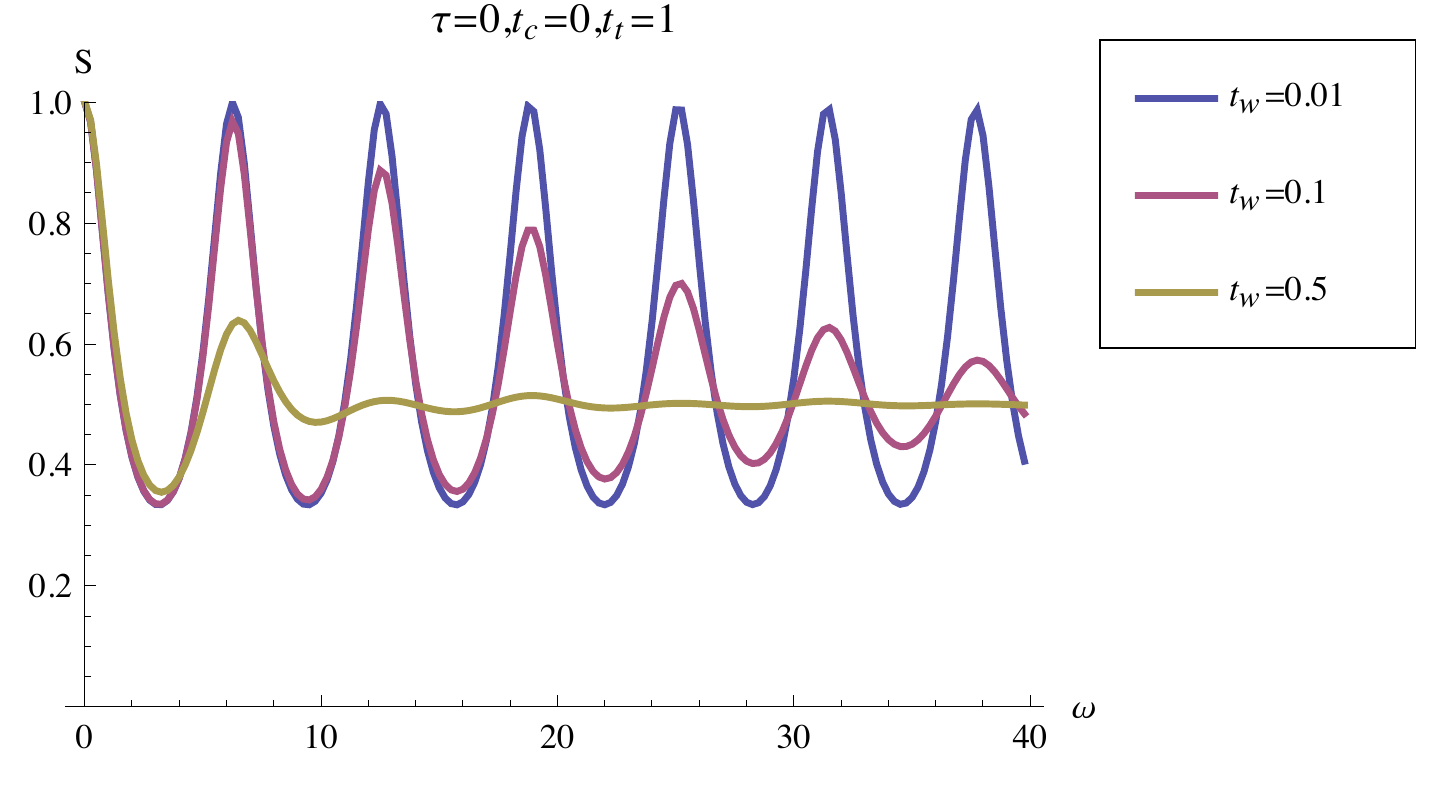}
\includegraphics[width=\linewidth]{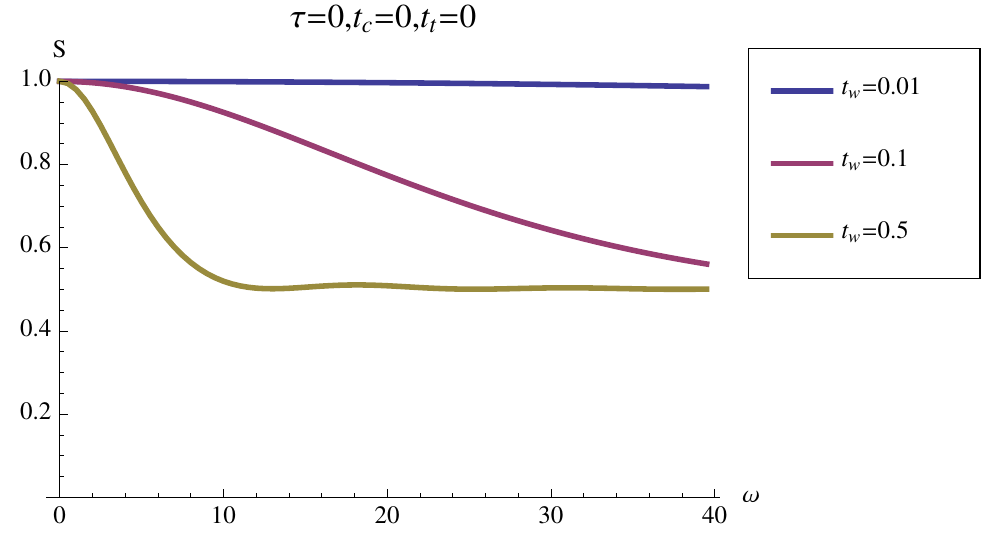}
\caption{(Color online) Truth table similarity with no time shift, against frequency shift. i.e. we have pure frequency mismatch and no temporal mismatch. (top) Time-resolved detection, where $t_c=0$ and $t_t=1$. (bottom) Gated detection, where $t_c=t_t=0$. The different lines correspond to different detector integration windows $t_w$.} \label{fig:freq_mis}
\end{figure}

In Fig. \ref{fig:freq_mis} we consider the case where there is only frequency mismatch and no temporal mismatch. We consider two separate cases -- gated detection and time-resolved detection. For time-resolved detection (top) the similarity against frequency exhibits oscillatory behaviour, where we have chosen the click times arbitrarily to illustrate the general nature of the dynamics (a discussion on the effects of click times will be presented later). For narrow detector integration times the oscillations periodically approach perfect similarity, and as the integration time increases the oscillations become damped. Thus for narrow integration times there are periodic frequencies at which perfect gate operation can be attained, whereas for larger detection windows perfect gate operation can only be achieved when there is no frequency mismatch. In the case of gated detection (bottom), our detectors are only `open' for a short interval, so we enforce $t_c=t_t=0$. In this case perfect gate operation is possible, provided that the integration time is short. For larger integration times the similarity decreases monotonically. The same results are observed when the gate operates in the diagonal basis, demonstrating that the behaviour of the gate is truly quantum mechanical.

\begin{figure}[h!tb]
\includegraphics[width=\linewidth]{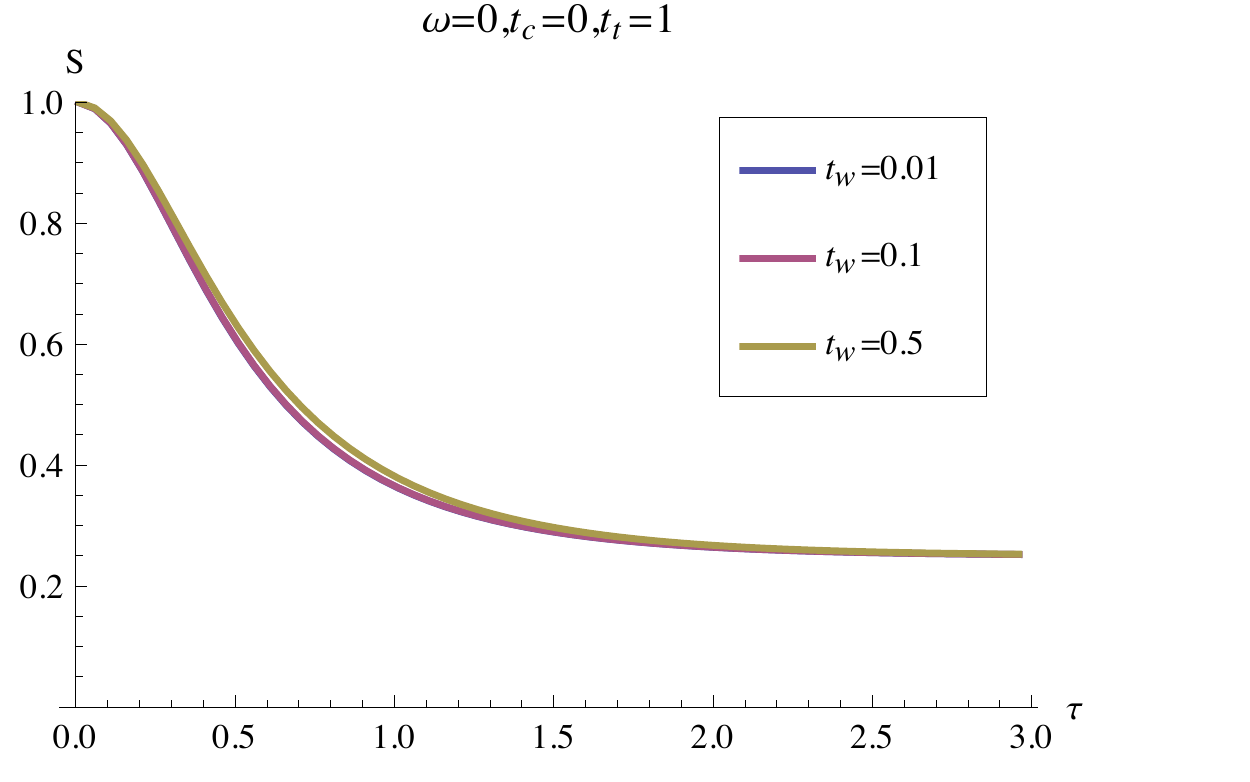}
\includegraphics[width=\linewidth]{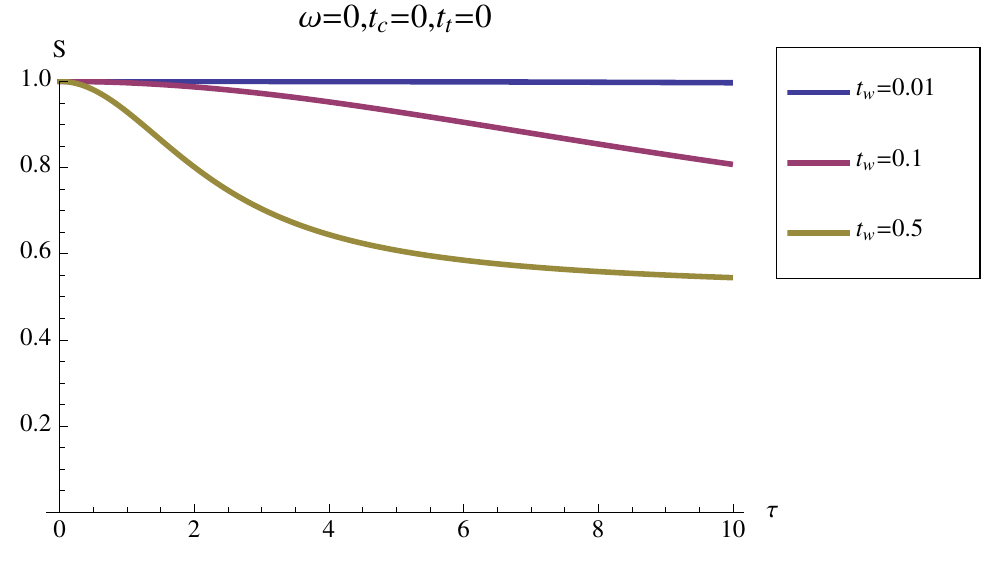}
\caption{(Color online) Truth table similarity with no frequency shift, against temporal shift. i.e. we have pure temporal mismatch and no frequency mismatch. The different lines correspond to different detector integration windows $t_w$. (top) Time-resolved detection, $t_c=0$ and $t_t=1$. (bottom) Gated detection, where both photons are measured at time $t_c = t_t = 0$.} \label{fig:temp_mis}
\end{figure}

In Fig. \ref{fig:temp_mis} we consider the converse situation where there is only temporal mismatch and no frequency mismatch. Unlike the previous situation there is no oscillatory behaviour and the similarity decreases monotonically with the temporal mismatch, regardless of the integration time. In the case of gated detection it is possible to achieve perfect gate operation for small detector integration times. This is not the case for time-resolved detection since the click times reveal information about which photon is which.

Intuitively one would expect quantum behaviour within the gate to vanish with the introduction of mode-mismatch. We emphasise that we are not solely considering \emph{complete} mode-mismatch, but \emph{partial} mismatch. That is, each photon is characterised by a temporal distribution function, and we consider finite relative frequency and temporal translations between the photons. Thus, even with mode-mismatch, not \emph{all} quantum behaviour is lost. Only in the limits $\omega\to\infty$ or $\tau\to\infty$ (and in the absence of filtering) does quantum behaviour disappear entirely.

\begin{figure}[h!tb]
\includegraphics[width=\linewidth]{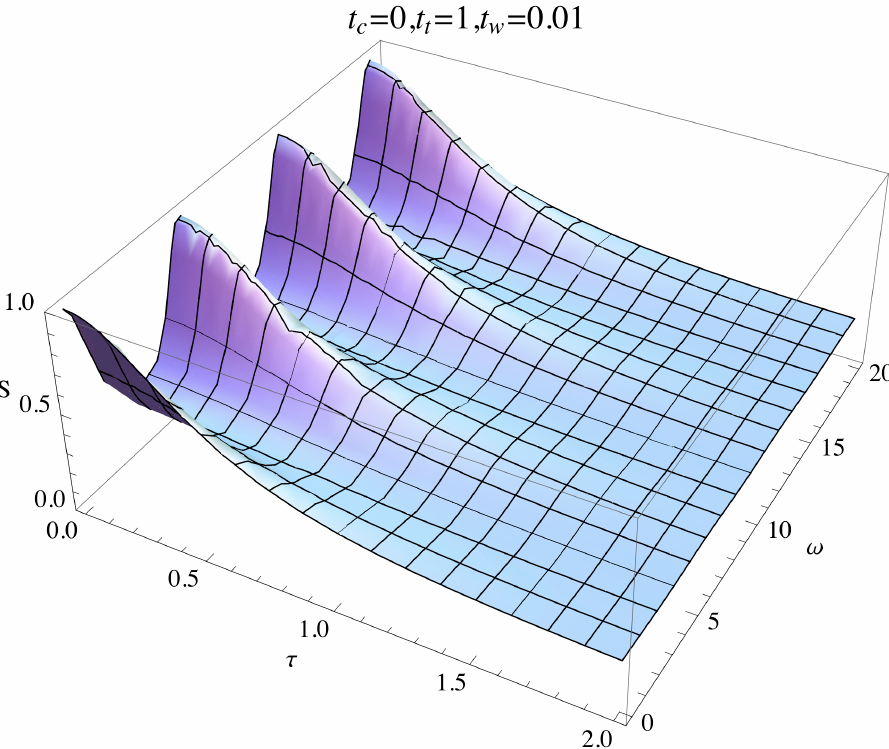}
\caption{(Color online) Truth table similarity against time shift and frequency shift, with a narrow detector integration time $t_w=0.01$. We have time-resolved detection, $t_c=0$ and $t_t=1$. When $t_c = t_t = 0$ and $t_w = 0.01$, $S\approx 1$ for all $\tau$ and $\omega$ (not shown).} \label{fig:temp_freq}
\end{figure}

In Fig. \ref{fig:temp_freq} we combine the previous two plots into a plot against both the temporal and frequency shifts. We set $t_c=0$ and $t_t=1$, i.e. time-resolved detection with two different click times. On the $\omega$ axis we observe the oscillations as before, while on the $\tau$ axis we observe the monotonic decrease in the similarity, without oscillations. With gated detection and a short integration time (not shown in Fig. \ref{fig:temp_freq}) we observe perfect gate operation for all $\tau$ and $\omega$, i.e. the filtering protects us against distinguishing information between the photons. This is an important observation as it implies that imperfect photon preparation can be overcome with the use of appropriate detectors and filtering. This is not surprising to present day experimentalists, who routinely employ narrowband filtering to improve the fidelity of their gates, albeit at the expense of gate success probability.

\begin{figure}[h!tb]
\includegraphics[width=\linewidth]{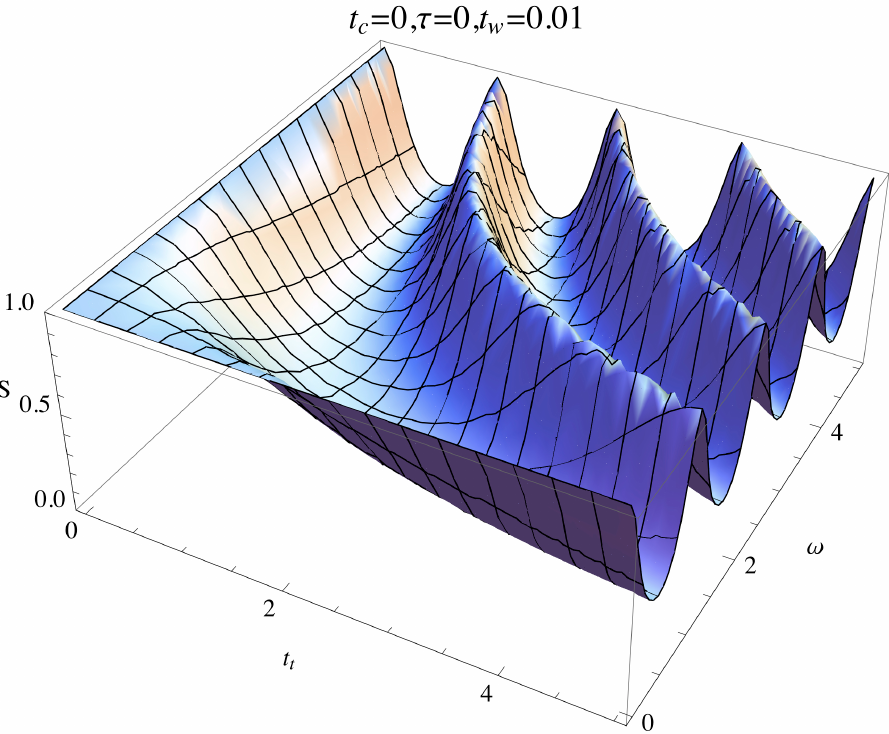}
\caption{(Color online) Truth table similarity against frequency shift and target photon detection time $t_t$. The control photon is detected at $t_c=0$. A narrow integration window is used, $t_w=0.01$, and there is no time shift $\tau=0$.} \label{fig:freq_det}
\end{figure}

Finally, in Fig. \ref{fig:freq_det} we consider the behaviour of the gate as a function of the detection time of the target photon. We have set $t_c=0$, no temporal shift, $\tau=0$, and a narrow integration time, $t_w=0.01$. When $\omega=0$ the gate behaves perfectly, since this corresponds to the situation where the photons are completely indistinguishable in both frequency and time. Additionally, when $t_t=0$ we also observe perfect gate operation, since now \mbox{$t_c=t_t=0$} so the detection events do not reveal any distinguishing information about the two photons. In the intermediate case, the similarity of the gate's operation oscillates against both the target's detection time $t_t$ and the frequency shift $\omega$. Depending on the frequency mismatch there are multiple values for detection time which result in ideal gate operation. Note that the frequency of the oscillations of $S$ against $\omega$ increases with the difference between detection times. A similar oscillatory behaviour against the difference in detector click times for the case of an an entangling operation between atomic qubits was observed in Ref. \cite{bib:MetzBarrett08}.

It is evident that temporal and frequency mismatch exhibit quite distinct properties. Temporal mismatch generally results in a monotonic deterioration against $\tau$, whereas frequency mismatch exhibits oscillatory behaviour against $\omega$. The question arises as to how this asymmetry comes about, since time and frequency are conjugate variables. The symmetry is broken as a result of the detector model, which does not implement the same operation in time-space as it does in frequency-space. We expect that with different detector models, e.g. frequency-resolved detection, the nature of these observations would differ. Presumably, with frequency-resolving detection rather than time-resolving detection, the role of frequency and time would be reversed in the presented results. The oscillatory behaviour against frequency mismatch is perhaps surprising. The intuition here is that a frequency shift induces a complex rotation factor in the time-domain giving rise to oscillations.

Given that gating appears to be an inevitable requirement for high-fidelity gate operation (which is easily physically implemented and already widely employed), strategies must be adopted to deal with gate non-determinism, this being the side-effect of gating. Many authors have given consideration to approaches for dealing with gate failure in an efficient manner, most notably using cluster-state \cite{bib:Raussendorf01,bib:Raussendorf03} approaches, which have been shown to allow efficient computation in the presence of gate failure \cite{bib:Nielsen04,bib:BrowneRudolph05}.

In conclusion, we have examined the difference between temporal and frequency mismatch in an elementary LO quantum gate. We demonstrated that the assumptions about the detector model have a strong impact on the operation of the gate. In general, with a gated detector model and small detector integration times, perfect gate operation can be achieved, while with other detector models the fidelity of the gate deteriorates. As photons produced for LO protocols are typically independently prepared, understanding the effects of frequency and temporal mismatch is valuable, and understanding the spectral properties of photo-detectors and their impact on the gate is crucial.

\begin{acknowledgments}
This research was conducted by the Australian Research Council Centre of Excellence for Quantum Computation and Communication Technology (Project number CE110001027). We thank Christine Silberhorn for helpful discussions.
\end{acknowledgments}

\bibliography{paper}

\end{document}